%% file: main.tex
\documentclass[sigconf]{acmart}
\AtBeginDocument{%
  }

\copyrightyear{2025}
\acmYear{2025}
\setcopyright{cc}
\setcctype{by}
\acmConference[ASSETS '25]{The 27th International ACM SIGACCESS Conference
on Computers and Accessibility}{October 26--29, 2025}{Denver, CO, USA}
\acmBooktitle{The 27th International ACM SIGACCESS Conference on Computers
and Accessibility (ASSETS '25), October 26--29, 2025, Denver, CO, USA}
\acmDOI{10.1145/3663547.3746362}
\acmISBN{979-8-4007-0676-9/2025/10}

\usepackage{listings}
\usepackage{xspace}
\usepackage{multirow}

\usepackage{xparse}

\usepackage{alltt}

\NewDocumentCommand{\codeword}{v}{%
\texttt{\textcolor{black}{#1}}%
}

\begin{document}

\title[A11yShape: AI-Assisted 3-D Modeling for Blind and Low-Vision Programmers]{A11yShape: AI-Assisted 3-D Modeling for \\ Blind and Low-Vision Programmers}

\author{Zhuohao (Jerry) Zhang}
\orcid{0000-0001-8708-1429}
\affiliation{%
  \institution{University of Washington}
  \city{Seattle}
  \state{Washington}
  \country{United States}
}

\author{Haichang Li}
\orcid{0009-0006-0952-0709}
\affiliation{%
  \institution{Purdue University}
  \city{West Lafayette}
  \state{Indiana}
  \country{United States}
}

\author{Chun Meng Yu}
\orcid{0009-0000-1324-044X}
\affiliation{%
  \institution{Purdue University}
  \city{West Lafayette}
  \state{Indiana}
  \country{United States}
}

\author{Faraz Faruqi}
\orcid{0000-0002-1691-2093}
\affiliation{%
  \institution{MIT CSAIL}
  \city{Cambridge}
  \state{Massachusetts}
  \country{United States}
}

\author{Junan Xie}
\orcid{0009-0006-8201-8563}
\affiliation{%
  \institution{The Hong Kong University of Science and Technology (Guangzhou)}
  \city{Guangzhou}
  \country{China}
}

\author{Gene S-H Kim}
\orcid{0000-0001-9514-4610}
\affiliation{%
  \institution{Stanford University}
  \city{Stanford}
  \state{California}
  \country{United States}
}

\author{Mingming Fan}
\orcid{0000-0002-0356-4712}
\affiliation{%
  \institution{The Hong Kong University of Science and Technology (Guangzhou)}
  \city{Guangzhou}
  \country{China}
}

\author{Angus Forbes}
\orcid{0000-0002-8700-7795}
\affiliation{%
  \institution{NVIDIA}
  \city{Santa Clara}
  \state{California}
  \country{United States}
}

\author{Jacob O. Wobbrock}
\orcid{0000-0003-3675-5491}
\affiliation{%
  \institution{University of Washington}
  \city{Seattle}
  \state{Washington}
  \country{United States}
}

\author{Anhong Guo}
\orcid{0000-0002-4447-7818}
\affiliation{%
  \institution{University of Michigan}
  \city{Ann Arbor}
  \state{Michigan}
  \country{United States}
}

\author{Liang He}
\orcid{0000-0003-4826-629X}
\affiliation{%
  \institution{University of Texas at Dallas}
  \city{Richardson}
  \state{Texas}
  \country{United States}
}

\renewcommand{\shortauthors}{Zhang et al.}

\input{sections/0-abstract}


\begin{teaserfigure}
    \centering
    \includegraphics[width=\textwidth]{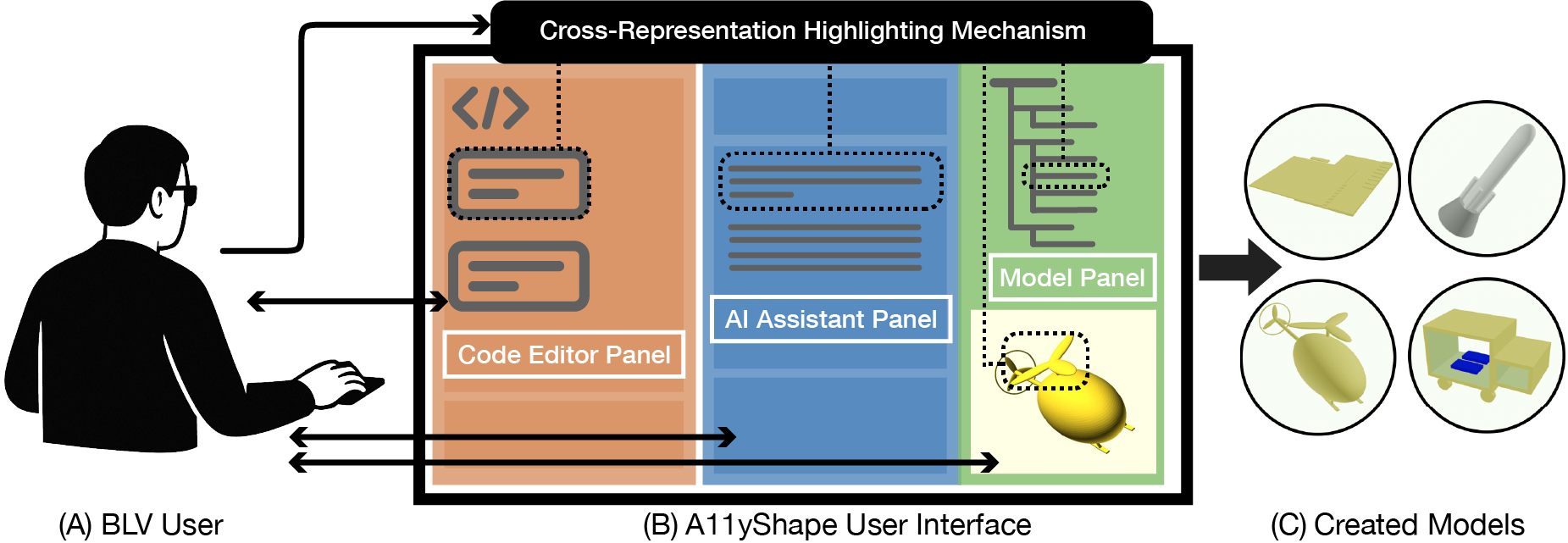}
    \caption{With A11yShape, (A) a blind or low-vision (BLV) user can create, interpret, and verify 3-D models through (B) a user interface composed of three parts: Code Editor Panel, AI Assistant Panel, and Model Panel. These panels are linked by a cross-representation highlighting mechanism that connects code, textual descriptions, hierarchical model abstractions, and 3-D visual renderings. The system supports the creation of (C) diverse, customized 3-D models created by BLV users.}
    \Description{A schematic overview of a blind or low-vision (BLV) user interacting with A11yShape to create 3-D models. Panel A shows the BLV user. Panel B illustrates the A11yShape interface, which includes three components: Code Editor panel, AI Assistant Panel providing textual descriptions, and Model Panel showing a hierarchical structure and 3-D renderings. A cross-representation highlighting mechanism links elements across these panels to support non-visual navigation in the user interface. Panel C displays examples of 3-D models created by the user.}
    \label{fig:teaser}
\end{teaserfigure}
\maketitle

\input{sections/1-intro}
\input{sections/2-related-work}

\input{sections/3-system}

\input{sections/4-study}
\input{sections/5-result}

\input{sections/6-discussion}
\input{sections/7-conclusion}

\bibliographystyle{ACM-Reference-Format}
\bibliography{sample-base}

\input{sections/99-appendix}

\end{document}

%% file: sections/0-abstract.tex
\begin{abstract}
      Building 3-D models is challenging for blind and low-vision (BLV) users due to the inherent complexity of 3-D models and the lack of support for non-visual interaction in existing tools. To address this issue, we introduce \textit{A11yShape}, a novel system designed to help BLV users who possess basic programming skills understand, modify, and iterate on 3-D models. A11yShape leverages LLMs and integrates with OpenSCAD, a popular open-source editor that generates 3-D models from code. Key functionalities of A11yShape include accessible descriptions of 3-D models, version control to track changes in models and code, and a hierarchical representation of model components. Most importantly, A11yShape employs a cross-representation highlighting mechanism to synchronize semantic selections across all model representations---code, semantic hierarchy, AI description, and 3-D rendering. We conducted a multi-session user study with four BLV programmers, where, after an initial tutorial session, participants independently completed 12 distinct models across two testing sessions, achieving results that aligned with their own satisfaction. The result demonstrates that participants were able to comprehend provided 3-D models, as well as independently create and modify 3-D models---tasks that were previously impossible without assistance from sighted individuals.
\end{abstract}

%% file: sections/1-intro.tex
\section{Introduction}

Recent advances in large language models (LLMs) have significantly expanded opportunities for blind and low-vision (BLV) individuals to independently perform creative tasks previously inaccessible without sighted assistance. With the emergence of readily available and cost-effective LLMs, BLV users now have greater potential to independently engage with visually complex tasks like image editing \cite{lee24altcanvas, chang24editscribe} or application programming \cite{saben2024enabling, zhang2025distinguishing}, potentially benefiting the 1.7\% of programmers with visual impairments \cite{StackOverflow2022} and others previously excluded from programming due to visual barriers \cite{mountapmbeme2022addressing, pandey2021understanding, albusays2017interviews}. 
Despite these promising developments, highly intricate visual-spatial workflows such as three-dimensional (3-D) modeling remain under-explored in accessibility research. 3-D modeling challenges even sighted users due to its demands on spatial reasoning, complex visual interfaces, tricky input articulations, and mental visualization requirements. In the meantime, programming interfaces for 3-D modeling may provide a more accessible pathway for BLV users, as text-based code can transform abstract spatial concepts into concrete, rule-based instructions that don't rely on visuals. For BLV programmers, designers, and students who need to understand and create 3-D models, investigating accessible approaches to 3-D modeling under these new possibilities is essential.

We introduce \textit{A11yShape}, an interactive 3-D modeling system developed through participatory design with a BLV co-author. Unlike traditional approaches that rely on author-curated captions for model descriptions, A11yShape leverages both the underlying model code and rendered images to produce detailed and accurate textual explanations of 3-D models, which have been validated through user studies to achieve high ratings in author-curated evaluation metrics. A11yShape integrates OpenSCAD \cite{OpenSCAD}, a commonly used code-to-model environment, with advanced capabilities of GPT-4o to synthesize complementary textual descriptions. Furthermore, A11yShape extends interactive access through three attributes: (1) a hierarchical representation enabling structured navigation of model components, (2) integrated version control to track iterative model changes, and (3) an interactive verification loop allowing BLV users to directly query and validate spatial attributes or design decisions. Most importantly, A11yShape employs a dynamic cross-representation highlighting mechanism that synchronizes selections across multiple model representations, enabling users to seamlessly navigate among code, semantic hierarchy, AI-generated descriptions, and rendered elements. 

To investigate how BLV programmers engage with A11yShape, we conducted a multi-session study involving four BLV participants each completing three separate, successive sessions, totaling 12 sessions. We observed that, although with notable flaws like components being misaligned or overlapping in conflict, participants successfully performed previously inaccessible tasks to create three complete 3-D models both by guidance and in free-form over the sessions. After the initial tutorial session, participants independently created 12 distinct models across the testing sessions, with outcomes that met their own satisfaction. To encourage deeper engagement with the system, participants were given ample time, each spending approximately four hours to complete three models. Our findings revealed that the cross-representation highlighting mechanism enabled fluid navigation between different ways of understanding models, and AI-generated descriptions in particular were perceived as compensating for the lack of visual verification. Participants developed distinctive workflows based on varying levels of AI assistance and adopted strategic approaches including: incremental building through AI-verification loops, leveraging semantic hierarchies for error correction, and using real-world metaphors for mental model construction. Despite overall success, certain challenges emerged, including high cognitive load from interpreting textual descriptions, difficulty in understanding spatial relationships, and uncertainty about operation success in the absence of visual or tactile feedback. These findings suggest promising directions for assistive technologies that can empower BLV users to independently engage with inherently visual creative workflows.

In summary, this paper makes the following contributions:
\begin{itemize}
\item A11yShape, the first AI-assisted 3-D modeling system leveraging code-augmented LLM descriptions, hierarchical component navigation, and interactive verification loops.
\item Empirical insights from an extensive multi-session user study that reveals how BLV users navigate spatial cognition challenges, develop mental models, and employ different strategies to create desired 3-D models without visual feedback.
\end{itemize}

%% file: sections/2-related-work.tex
\section{Related Work}
Our work is built upon prior work in three key areas: the use of LLMs and other AI techniques for generating and interpreting 3-D models, approaches that explore modalities for supporting BLV users in 3-D modeling and understanding, and AI-assisted tools that support creativity tasks for BLV users. 

\subsection{AI-Driven 3-D Model Generation}
Recent work has explored the use of generative models and large language models for automatic 3-D content creation in various applications, such as avatar and scene generation \cite{text23dsurvey} and modifiable 3-D model asset generation \cite{3D-gpt,siddiqui2024assetgen}. Text-to-3-D systems, such as \textit{DreamFusion} \cite{poole2022dreamfusion}, \textit{Magic3D} \cite{magic3D}, and \textit{Text2Mesh} \cite{text2mesh}, allow users to create desired 3-D geometries from natural language prompts. These systems leverage pretrained 2D visual-language models as guidance to synthesize geometry and texture from textual input. Besides the methods for 3-D generation directly from natural language, researchers have explored combining textual prompts with visual knowledge, like images, to generate more grounded and view-consistent 3-D content \cite{vp3D,dream3D, luciddreamer}. For example, \textit{DreamBooth3D} \cite{dreambooth3D} allows users to generate personalized 3-D models of a specific object or person based on a few reference images. 
While these methods support the 3-D generation without needing domain-specific training or 3-D datasets, it is difficult for users, especially BLV people, to interpret and validate the generated 3-D content. 
To address this, researchers grounded language in 3-D representations to interpret and interact with AI-generated 3-D models \cite{kestrel,shapellm,chen2018text2shape}. For example, \textit{Cap3D} \cite{cap3d} generates descriptive captions of 3-D objects using pre-trained models in image captioning, image-text alignment, and LLMs to consolidate information from multiple rendered views, providing textual explanations of the generated 3-D content for understanding. These systems remain limited in offering actionable feedback about a 3-D model’s structure, orientation, or completeness—critical aspects for users, especially BLV people, to validate the generated 3-D shape. Our system aims to provide textual descriptions that summarize the generated model with detailed information, while offering actionable aids to further edit it. 

While AI shows promise in generating 3-D models, human creativity in crafting models still takes an irreplaceable role \cite{bebeshko20213D}. Researchers have also explored human-AI collaborative approaches to facilitate 3-D modeling \cite{3Ddescription}. For example, 3\textit{DALL-E} \cite{3dall-e} integrates text-to-image diffusion models into 3-D modeling workflows and uses image generation as a semantic design aid to facilitate shape design. \textit{Style2Fab} \cite{style2fab} enables users to personalize 3-D models using generative AI while preserving their functionality. Our work extends this body of research and explores how AI can assist BLV users in authoring, interpreting, and confirming custom 3-D models through programming and descriptive feedback. 

\subsection{3-D Model Accessibility for BLV Users}
A large body of prior work has demonstrated the potential of making 3-D models accessible and interactive to support BLV users across a wide range of non-visual tasks \cite{shi_assets17, sargsyan_chi23, linespace, gotzelmann_taccess_18}, such as conversational interfaces \cite{reinders_dis_23, i3ms}, learning \cite{tickerstalkers, shi_chi19, Anđić07082024}, and orientation and mobility (O\&M) training using map-based representations \cite{tacticon, lucentmaps, onbtraining, accessmaps, leporini_taccess20}. In other contexts, accessible 3-D models have enabled tangible interaction with circuits \cite{accessiblecircuits, tangiblecircuits} and enriched storytelling experiences in tactile books~\cite{kimtactilebook}. Despite this potential, most accessible 3-D models are still created by sighted people or using computer vision-based approaches \cite{feliceIEEE2005, Schwarzbach2012}, as BLV users continue to face barriers in creating custom 3-D models. While AI-assisted approaches have enabled the rapid creation of 3-D content, researchers have explored modalities and techniques for 3-D structure interpretation and modeling \cite{daedlus_uist21, Rossetti_2018}. Touch and auditory feedback are the most common modalities. For example, \textit{TouchPilot} \cite{touchpilot} provides step-by-step audio guidance to assist blind users in exploring and understanding complex 3-D structures. Lieb \textit{et al.} developed an audio-haptic system that enables blind users to independently inspect and verify 3-D models created through text-based modeling tools like OpenSCAD \cite{lieb_assets20}. Recently, shape-changing displays have also been used to support 3-D modeling for BLV people. For example, \textit{shapeCAD} \cite{shapeCAD} presents an accessible 3-D modeling workflow through a 2.5D tactile shape display that uses actuated pins to render the physical shapes of digital 3-D models, enabling blind and visually impaired users to explore and iteratively refine their models in real-time. In contrast, our system introduces a low-cost, software-based approach that allows BLV users to programmatically create, interpret, and verify 3-D models, leveraging the power of AI. 

\subsection{AI-Assisted Creativity and Creativity Accessibility for BLV Users}

While much of accessibility research has focused on enabling BLV users to read and consume digital content, recent efforts have increasingly shifted toward empowering BLV users as content creators \cite{ladner2015design}. In particular, creativity support has emerged as a key focus in accessibility research, with a growing body of work designing tools to support creative expression across a range of media.

Researchers have explored assistive technologies for various creative domains: document editing \cite{lee22collabally, das2022co11ab, das2022design}, photography \cite{hirabayashi23visphoto, adams16vizsnap}, image editing and generation \cite{huh2023genassist, chang24editscribe, lee24altcanvas}, emoji composition \cite{zhang21voicemoji}, programming \cite{herskovitz24programa11y, mountapmbeme22accessibleblockly, potluri22codewalk, mountapmbeme2022addressing, seo2023coding, albusays2017interviews, pandey2021understanding, saben2024enabling, zhang2025distinguishing, StackOverflow2022, mowar2025codea11ymakingaicoding}, presentation slides \cite{pengchi2023slidegestalt, zhang23a11yboard, zhang23accessibleslide, peng21sayitall}, website design \cite{li19spatial, li22tangiblegrid, potluri19multimodal, designchecker}, video scripting and editing \cite{huh2023avscript}, and other forms of media \cite{chhedakothary25artinsight, johnstone2024, potluri22psst, herskovitz23Diyats, saha23tutoria11y, zhang2024charta11y}.

These systems adopt both traditional multimodal interaction paradigms and, more recently, human-AI collaboration approaches to enhance accessibility in creative tasks \cite{designchecker, huh2023avscript, huh2023genassist, chang24editscribe, lee24altcanvas, mowar2025codea11ymakingaicoding, chhedakothary25artinsight}. As generative AI continues to improve, it opens new opportunities for making creative domains that were previously considered inaccessible such as 3-D model creation more approachable for BLV users. Designing 3-D content typically requires spatial reasoning and visual feedback, often mediated through tactile representations or external assistance for BLV users.

With A11yShape, we build on these prior works by extending the possibility of accessible creativity to 3-D modeling, a domain that has seen limited exploration in the context of BLV accessibility. Our system design is informed by the interaction techniques and AI-assisted paradigms introduced in this growing body of creativity-support literature. For example, we drew from how \textit{EditScribe} \cite{chang24editscribe} assist BLV users to make image edits through AI verification loops.

%% file: sections/3-system.tex
\section{The A11yShape System}

\begin{figure*}[h]
    \centering
    \includegraphics[width=1\textwidth]{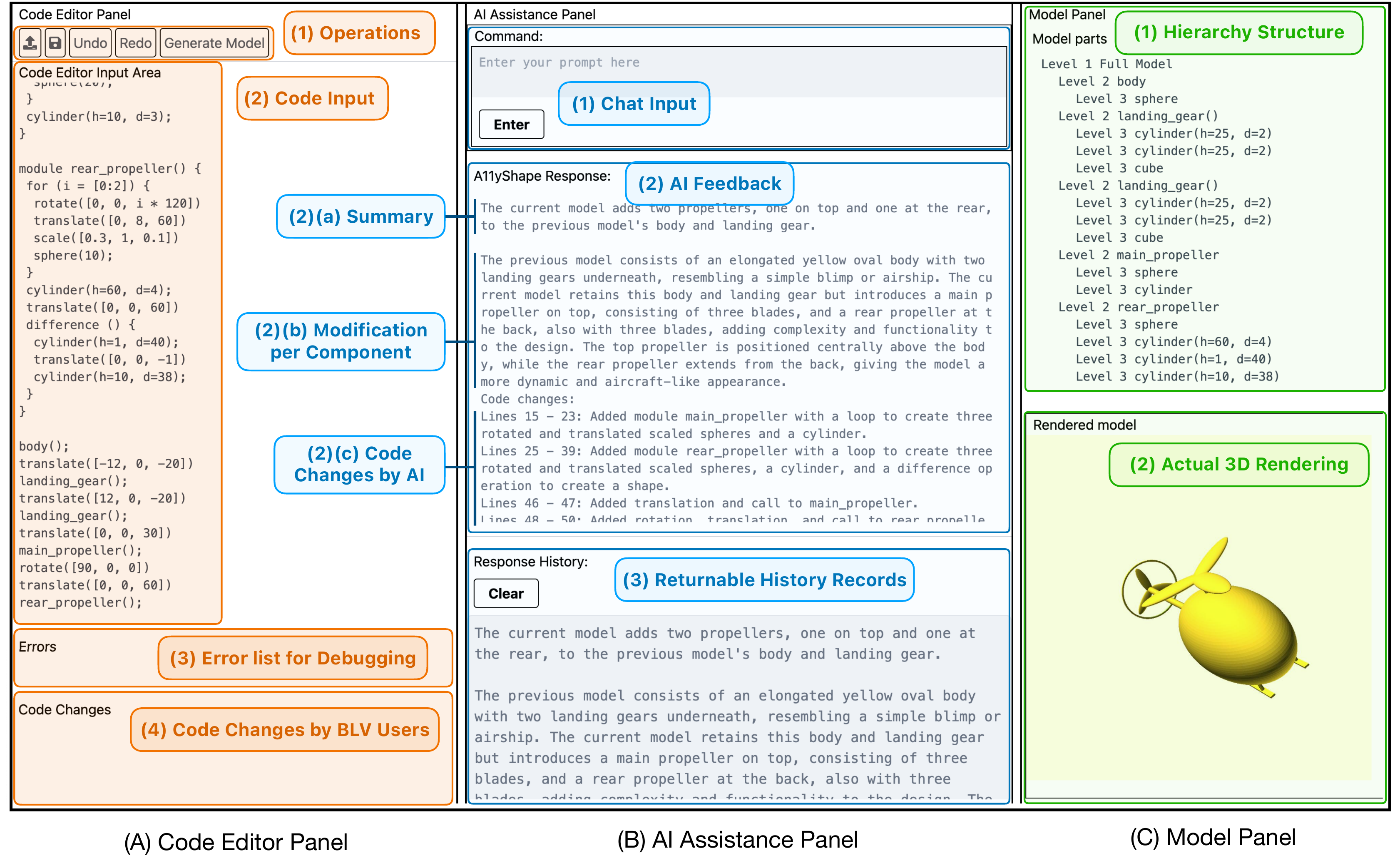}
    \caption{The A11yShape web interface for accessible 3-D modeling featuring: (A) the Code Editor Panel with programming capabilities, (B) the AI Assistance Panel providing contextual feedback, and (C) the Model Panel displaying hierarchical structure and 3-D rendering of the resulting helicopter model with propellers.}
    \label{fig:ui}
    \Description{Interface layout of A11yShape, an accessible 3-D modeling tool with three panels. Panel A (Code Editor) shows the programming interface with operation buttons and the code input area. Panel B (AI Assistance) displays contextual feedback with summarized model changes. Panel C (Model Panel) presents the hierarchical structure of components and a 3-D rendering of a yellow helicopter object with propellers. The interface demonstrates how blind and low vision users can create 3-D models through code with AI assistance.}
\end{figure*}

We present \textit{A11yShape}, an accessible 3-D modeling system developed through participatory design. A11yShape leverages a dynamic \textit{cross-representation highlighting} mechanism at its core to support BLV users to understand and edit 3-D models through all model representations including code, semantic hierarchy, AI descriptions, and 3-D rendering projections. We first present the A11yShape system, including its components and the aforementioned cross-representation highlighting mechanism. We then describe a real user journey drawn directly from one of our study sessions to illustrate how A11yShape supports BLV users in practice. We also present the details of our participatory design process.

\subsection{System Overview}
Built on a free code-to-model software OpenSCAD, A11yShape introduces a four-facet representation of 3-D models through connecting the source code, a hierarchical model abstraction, AI-generated textual descriptions, and the actual rendering of the model. This architecture allows users to explore models, query design properties, and apply modifications in a dynamic representation. Users can write codes in the editor, easily navigates to chat input window for asking questions or making edits, and traverse through the hierarchical abstraction for more structured access.
A11yShape is implemented as a web-based interface using Python's Flask web framework. It uses GPT-4o (\codeword{gpt-4o-2024-08-06}) \cite{gpt4o} for generating detailed textual descriptions and managing OpenSCAD model edits. We plan to open source A11yShape\footnote{Project Github repository: https://github.com/DE4M-Lab/A11yShape} to the BLV community.

\subsection{User Interface Components}
A11yShape consists of three main interactive panels (Figure~\ref{fig:ui}): the Code Editor, the AI Assistance Panel, and the Model Panel, each serving distinct interaction purposes that collectively support accessible 3-D modeling.

\subsubsection{Code Editor}

The Code Editor (Figure~\ref{fig:ui}A) provides BLV users a fully accessible, standard text-editing interface optimized for OpenSCAD code. Essential functionality includes rendering the model from written code, uploading existing files, saving progress, and quick access to debugging through an integrated \textit{Error Log}. This accessible debugging feature explicitly highlights syntax that prevents successful renders, thus streamlining error correction during iterative model modifications. Additionally, a dedicated \textit{Code Changes List} displays code alterations after manual modifications, allowing tracking the recent changes.
For example, when the user changes a cylinder component's height and diameter, the list will be populated with a new record: ``Line 22: Cylinder's parameter (height, diameter) changed, from (h=5, d=2) to (h=10, d=3).'' The change record is generated by prompting a separate LLM (Appendix \ref{sec:prompts_code}) to compare two versions of the code. Since the comparison focuses on small, localized structural differences in parameter values or geometry descriptions, this task is well-bounded and deterministic, making LLM-based summaries reliable for this purpose.

In addition, we provide basic keyboard shortcuts for programming operations, like uploading files, saving code, and navigating between different panels.

\subsubsection{AI Assistance Panel}

The AI Assistance Panel (Figure~\ref{fig:ui}B) offers a suite of tools for model understanding, creation, and version management. It consists of three integrated components: an always-available input chat box, an AI feedback panel, and a history record panel. Together, these components serve two core functions: an \textit{AI verification loop} for model understanding and a robust \textit{Version Control} system.

The \textit{AI verification loop} begins with the input chat box where users can request information about their current model or solicit modifications through natural language (prompts in Appendix \ref{sec:prompt_chat_input}). The system uses a multimodal prompting architecture that combines the model's or the current selected components' modular code, rendered images from multiple angles (generated using different camera views with the OpenSCAD engine), and user queries (Appendix \ref{sec:prompts_describing}). The rendered images are from the top, bottom, front, rear, left, and right side of the model to give LLMs a clear picture of the model to generate better descriptions. Together, the multimodal contexts were provided for the LLM to generate detailed and accurate responses. More specifically, the AI feedback panel structures LLM's responses into three distinct sections: (1) a concise \textit{summary} of changes, (2) a detailed description of \textit{modifications per component}, and (3) a \textit{code change list}.

\paragraph{AI Validation Study}
To evaluate whether the AI-generated descriptions on the models were accurate using different camera views and model codes, we conducted a validation study to test whether AI-generated descriptions were accurate enough for sighted users. We recruited 15 participants to rate eight generated model descriptions together with rendered projection images from the model's six principal views. Participants were asked to rate each description based on a set of key metrics using a 5-point Likert scale (1 = Poor, 5 = Excellent). We developed this metric set through internal discussions because there is no existing usable metric to evaluate AI-generated descriptions of 3-D models. 

\begin{itemize}
    \item \textit{Geometric Accuracy (M1)}: How accurately the description captures the geometric elements of \textit{individual components}, including shapes, proportions, and details?
    \item \textit{Spatial Relationships (M2)}: How well the description communicates the positioning and connections between different components, including relative locations and how parts fit together?
    \item \textit{Clarity \& Comprehensibility (M3)}: How well-written, organized, and understandable the description is, with appropriate use of language and logical flow?
    \item \textit{Completeness (M4)}: Whether the description includes all significant features visible across the 6 principal views without omitting important elements. This metric is related to ``false negatives'' of AI recognizing model components?
    \item \textit{Avoidance of Hallucinations (M5)}: Whether the description adheres strictly to what is actually present in the model without fabricating non-existent elements. This metric is related to ``false positives'' of AI recognizing model components?
\end{itemize}

The 15 participants were recruited through a physical computing course where students learned 3-D modeling and printing. They came from diverse academic backgrounds, including user experience design (7), industrial design (3), art (2), and computer science (3). All participants had hands-on experience with 3-D modeling for purposes such as animation, product design, and 3-D printing.
We found that the AI-generated model descriptions are reliable based on our results, as shown below. 

\begin{figure}[h]
    \centering
    \includegraphics[width=1\linewidth]{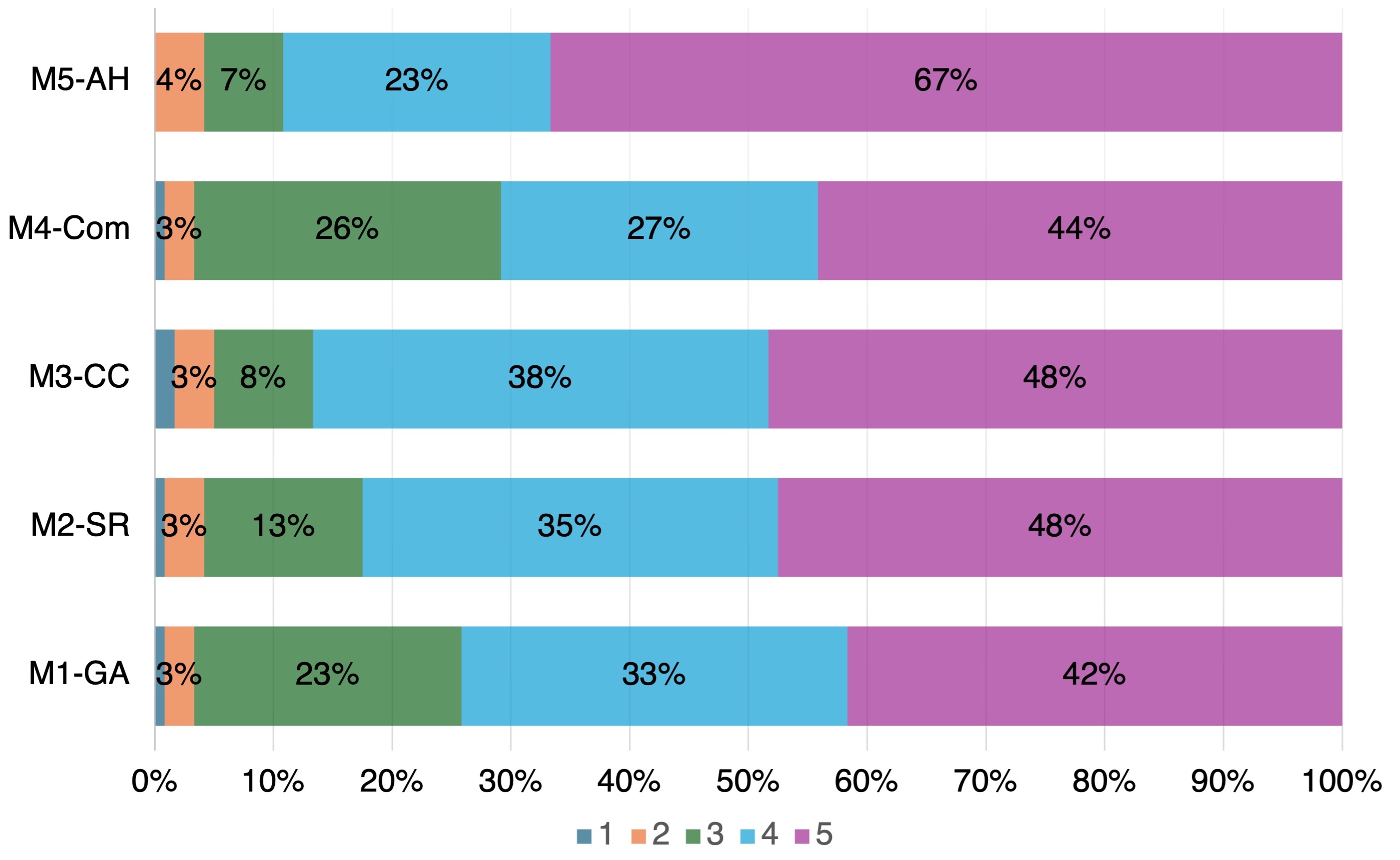}
    \caption{Results of the validation study, showing the Likert-scale distributions across five metrics. Most responses rated AI responses on models to be Good or Excellent.}
    \label{fig:validation_study}
    \Description{Stacked bar chart showing participant ratings (1 to 5 on a Likert scale) for five AI-generated 3D model descriptions (M1–M5). Each bar represents a different model and is segmented by rating percentages. The highest proportion for all models is in the "5 = Excellent" category, with M5-AH receiving the highest (66.7\%), followed by M3-CC (48.3\%), M2-SR (47.5\%), M4-Com (44.2\%), and M1-GA (41.7\%). Lower ratings (1 and 2) account for less than 10\% across all models, indicating overall positive evaluations. Colors correspond to ratings: 1 (blue), 2 (orange), 3 (green), 4 (light blue), and 5 (purple).}
    \vspace{-2mm}
\end{figure}

The strong performance across all metrics (with scores ranging from 4.11 to 4.52 out of 5.00) indicates that the AI-generated descriptions effectively capture the essential characteristics of 3-D models. Notably, the system demonstrated particular strength in avoiding hallucinations ($mean=4.52$, $SD=0.74$), suggesting high reliability in representing only features actually present in the models. Other results include spatial relationships ($mean=4.25$, $SD=0.84$) and clarity \& comprehensibility ($mean=4.28$, $SD=0.86$). The slightly lower but still strong scores in Completeness ($mean=4.11$, $SD=0.91$) and Geometric Accuracy ($mean=4.12$, $SD=0.86$) identify potential areas for future improvement, though all metrics achieved scores above 4.0, indicating overall excellent performance. We show the distribution of the Likert-scale scores in Figure \ref{fig:validation_study}. These findings validate our approach of using multimodal prompting with multiple camera views and modular code to generate model descriptions.

\begin{figure*}[h]
    \centering
    \includegraphics[width=1\textwidth]{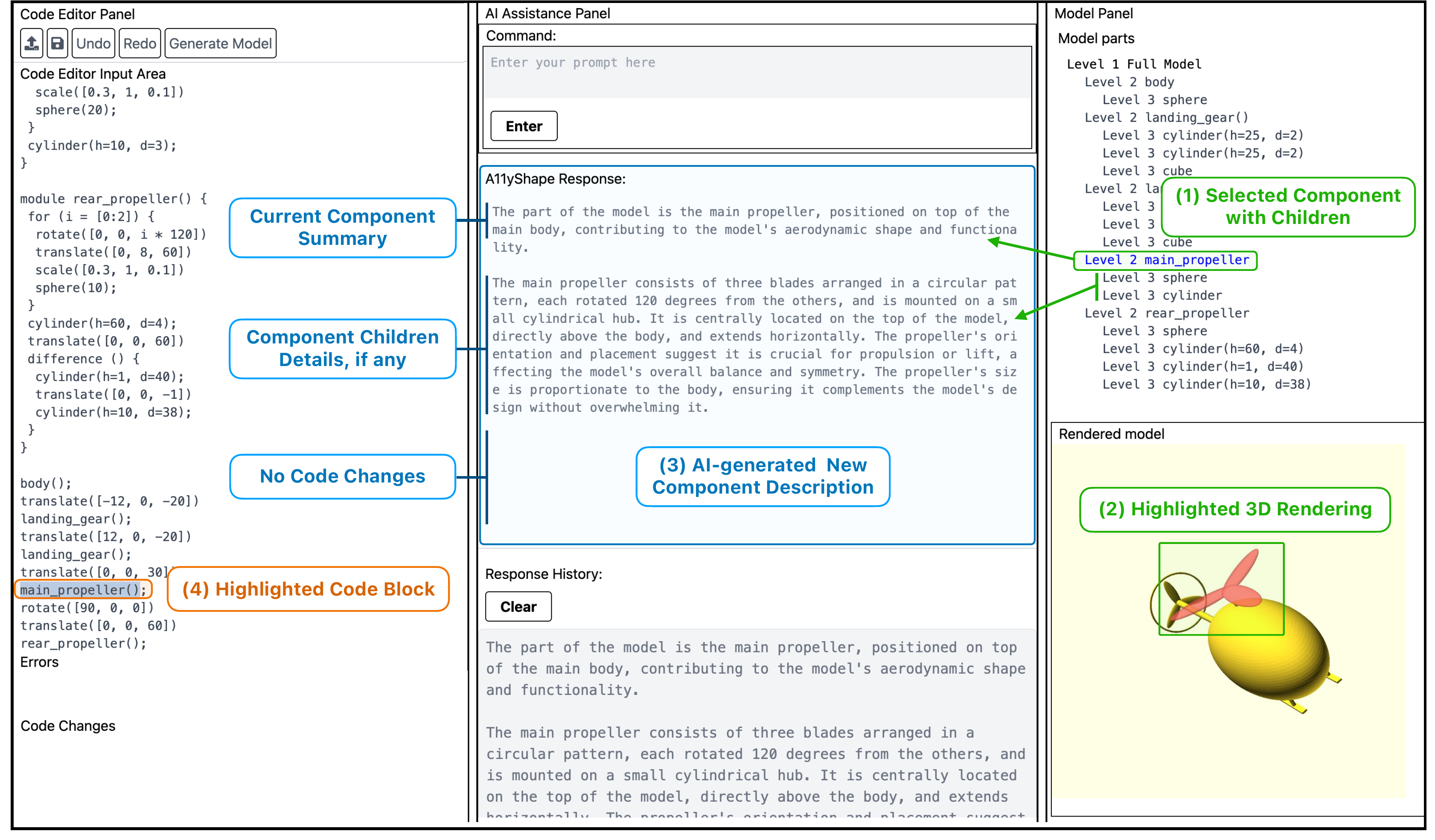}
    \caption{Dynamic cross-representation highlighting in A11yShape showing how the selection of the main propeller component in the hierarchical structure (1) automatically highlights corresponding elements in the 3-D rendering (2), triggers an AI-generated component description (3), and highlights the relevant code block (4).}
    \label{fig:core_interaction}
    \Description{A11yShape interface demonstrating dynamic cross-representation highlighting. The image shows how selecting the main_propeller component in the hierarchical tree structure (right panel) automatically synchronizes highlighting across all representations: the propeller appears highlighted in pink in the 3-D rendering of the yellow helicopter model, detailed semantic description is provided by the AI in the middle panel, and the corresponding code block is highlighted in the left panel's code editor. This synchronized multi-modal feedback enables blind and low-vision users to understand 3-D components through complementary representations.}
\end{figure*}

AI-generated descriptions also take real-world metaphors into account to help BLV users build imaginations of what the models roughly look like. Unlike the code changes list in the Code Editor Panel, which tracks manual user modifications, this list specifically documents AI-directed changes. By providing information of summary, change per component, and change per code, A11yShape enables BLV users to access different versions in a multi-representation format. To use the AI verification loop, when users select a code block, the panel automatically provides specific explanations of that code segment. Similarly, when a component is selected from the hierarchical tree, the panel displays a detailed narration describing that specific component's shape, position, and relationship to other parts. Furthermore, as users interact with A11yShape over time, all the endpoints are saved chronologically in the history record panel. Each history record in this panel is clickable, allowing users to read records and restore to an endpoint.

The code change list in both the code editor and the AI feedback panel, the history record panel, and the embedded undo/redo functionality serve together as the \textit{Version Control} system of A11yShape. The prompts we used to interpret codes (\textit{e.g.,} summarize code changes, compare models, \textit{etc.}) are attached in Appendix \ref{sec:prompts_code}.

\subsubsection{Model Panel}

The Model Panel (Figure~\ref{fig:ui}C) combines visual rendering with a hierarchical semantic structure, constituting the system's ``middle-layer'' among detailed code, AI-generated descriptions, and visual model output. The hierarchical list represents the model's internal structure by grouping OpenSCAD modules and primitives into nested semantic hierarchies. For example, in the helicopter example (Figure~\ref{fig:ui}C(1)), the two landing gears are at the same semantic level, each consisting of two cylindrical support legs connected by a long, flat rectangular base. They are grouped and listed semantically instead of in a plain list showing all model components. Users navigate through this tree structure, conceptually ``zooming in'' to inspect specific model components in detail. The 3-D rendering panel shows a model projection from a fixed angle that shows most parts of the model. It can be adjusted by using the chat input to indicate a specific angle (\textit{e.g.,} ``show the model from the top view''), or highlight a specific component (\textit{e.g.,} ``show the main propeller''). Then the LLM will generate a camera view parameter for the OpenSCAD engine to generate a corresponding image. Additionally, we offer a keyboard shortcut to switch between six orthographic views (ctrl+shift+number~1-6) of the top, bottom, front, rear, left, and right side of the model. When not specified, the image is rendered with a three-quarter view to show multiple sides of the model. Users can also switch to this default view by using a similar keystroke (ctrl+shift+number~0).

\subsection{Core Interaction: Cross-Representation Highlight}
When sighted people perform model constructing and editing, they use immediate visual feedback to verify model changes and match the changes to their mental model of expected changes \cite{hutchins1985direct}. However, such visual synchronization and mental model matching processes are not available to BLV users. Therefore, based on the introduced components, we further illustrate a dynamic cross-representation highlighting mechanism as A11yShape's core interaction to support accessible 3-D modeling. 

A11yShape maintains an active semantic synchronization across multiple representations: (1) OpenSCAD source code, (2) semantic hierarchical structure, (3) AI-generated textual descriptions, and (4) the actual visual renderings of the model. Upon any user selection in the code or hierarchical structure, A11yShape instantaneously highlights the corresponding component across all other representations, significantly reducing BLV users' cognitive load by simplifying the process of panel navigation, visual verification, and model editing. To notify screen reader users, A11yShape also uses audio feedback to indicate that the highlighting just happened and users could access different representations for better understanding. For example, when a user selects the propeller with three blades in the helicopter model from the hierarchical tree (Figure~\ref{fig:core_interaction}), the interface immediately (1) highlights the corresponding lines in the OpenSCAD code editor, (2) generates the specific description of the propeller in the AI feedback panel, and (3) visually emphasizes the exact rendered component in the model preview. In addition, when highlighting the rendered component in the preview image, the highlighted part will be colored differently and half-transparent to make sure it visually stands out and can show its connections and overlaps with other components.

Besides potential benefits like providing multiple levels of context simultaneously and alleviating cognitive load for blind users, this dynamic cross-representation highlighting is particularly designed for low-vision users by visually highlighting the component color in the rendered model area. 

\begin{figure*}[h]
    \centering
    \includegraphics[width=1\textwidth]{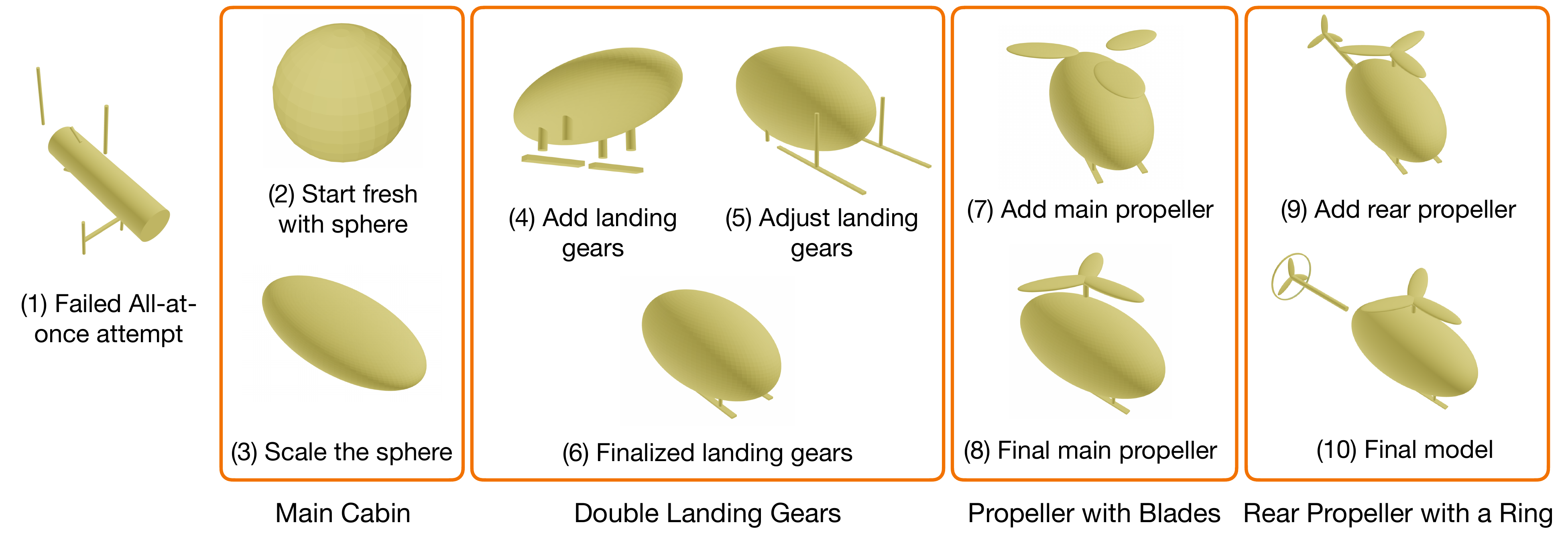}
    \caption{Incremental construction of a helicopter model in A11yShape showing: Main Cabin development from sphere to ellipsoid (1-3), Double Landing Gears addition and positioning (4-6), Propeller with Blades implementation (7-8), and Rear Propeller with Ring integration for the final model (9-10). The final model demonstrates some problems with component connecting.}
    \label{fig:user_journey}
    \Description{A step-by-step visualization of 3-D helicopter construction in the A11yShape environment. The image shows ten progressive stages organized in four columns: The first column shows a failed all-at-once attempt, followed by starting with a sphere and scaling it into an ellipsoid for the main cabin. The second column demonstrates adding landing gears and adjusting their positions. The third column shows the implementation of the main propeller with three blades. The fourth column displays the addition of a rear propeller with a protective ring, culminating in the final assembled helicopter model. Each stage is labeled with numbers 1-10 and brief descriptions of the construction steps.}
\end{figure*}

\subsection{User Journey}
We present an exemplary user journey of a low-vision programmer, ``Alex,'' who uses A11yShape to create a 3-D model of a helicopter. This scenario is from one of our study sessions, reflecting real interaction patterns. We illustrate his journey upfront to provide a clear picture of how A11yShape works. He began by attempting to create a complete helicopter model through a single detailed prompt. He described a helicopter with an elliptical body, dual landing gears, a three-bladed main propeller, and a tail propeller with specific orientations and positions for each component. His description of the model was based on his understanding of and prior searching on helicopter components. 

After the AI-generated model, Alex noticed discrepancies between his perceived result and the AI-generated result (Figure~\ref{fig:user_journey}(1)). He used his screen reader to navigate to and select the ``body'' component in the semantic hierarchy, which highlighted the rendered part, the corresponding code block, and triggered a verbal description: ``a long cylindrical body oriented along the y-axis.'' This showed that instead of his requested elliptical shape, the AI had created a cylinder. He continued exploring other components by selecting them in the hierarchy, discovering only one landing gear was visible despite requesting two, and the propellers' positions didn't match his specifications. Recognizing the limitations of his all-at-once approach, he decided to build the helicopter incrementally, component by component.

Alex began by crafting the body of the helicopter. He wrote code for a spherical shape (\codeword{sphere[50]}) and then used the system's description to understand its appearance. ``I need it more elliptical, not so round,'' he said after reading the AI's description. He adjusted parameters to stretch the sphere into an ellipsoid (Figure~\ref{fig:user_journey}(2-3)), repeatedly selecting the body component in the hierarchy to hear updated descriptions and verify the code changes. After several iterations of writing code, checking the description, and refining parameters, he achieved his desired elliptical body shape with a new scale (\codeword{scale([0.5, 1, 0.5]}).

With the body completed, Alex moved to construct the landing gear on top of the body. He added code for two symmetrical landing gears, each composed of cylinders and a flat cuboid. When selecting the landing gear component, the system description indicated two gears were present, but Alex's limited vision showed only one (Figure~\ref{fig:user_journey}(4)). This discrepancy caused momentary confusion: ``It sounds right in the description, but different from what I can barely see,'' he remarked. By examining the highlighted code, he realized both gears existed but weren't properly connected to the body. He adjusted the position parameters while repeatedly using the cross-representation highlighting to verify changes until the landing gear was properly attached to the elliptical body (Figure~\ref{fig:user_journey}(5)). The final code block for the landing gear is shown below (Figure~\ref{fig:user_journey}(6)):

{\small
\begin{verbatim}
    module landing_gear() {
       // First vertical support leg
       cylinder(h=25, d=2); 
       // 30 units forward
       translate([0, 30, 0])
       // Second vertical support leg
       cylinder(h=25, d=2);  
       translate([0, 10, 0])
       // Horizontal connecting base
       cube([3, 60, 1], center=true);  
    }
\end{verbatim}
}

For the main propeller implementation, Alex first created an empty function and prompted the AI to generate the propeller code. When the AI produced a three-blade design using a for-loop, he understood the intention to rotate each blade in the loop and selected it in the hierarchy to evaluate the result. ``The blades aren't meeting at the center point correctly,'' he noted after hearing the description. He attempted to fix it himself but wasn't satisfied with his modifications. Using the version control feature, he reverted to the AI's version and made targeted adjustments. By repeatedly selecting individual blades in the semantic hierarchy and hearing their positions described, he precisely adjusted each blade's angle and position until they formed a proper three-blade propeller configuration (Figure~\ref{fig:user_journey}(7-8)).

Finally, Alex added a decorative ring to the tail propeller. He created the initial ring and then selected it in the hierarchy to hear its dimensions. ``It needs to be larger to encircle the propeller blades,'' he determined. After a few size adjustments, each time selecting the component to verify its new dimensions through the verbal description, he achieved an appropriately sized ring that complemented the tail propeller. Furthermore, AI suggested that Alex could use OpenSCAD's boolean \codeword{difference()} function to create a more realistic propeller hub by subtracting a slightly smaller cylinder from the larger one, which would result in a thin circular ring that better represented the connection between the blades and the shaft. He took the suggestion and read both the code and newly generated AI description on the rear propeller to confirm the change (Figure~\ref{fig:user_journey}(9-10)). The code blocks for the top and rear propellers are finalized like this:

{\small
\begin{center}
\begin{minipage}[t]{0.48\textwidth}
\begin{alltt}
    module main_propeller() \{
        for (i = [0:2]) \{
            rotate([0, 0, i * 120])
            translate([0, 15, 10])
            scale([0.3, 1, 0.1])
            sphere(20);
        \}
        cylinder(h=10, d=3);
    \}
\end{alltt}
\end{minipage}
\hfill
\end{center}
}

\small{
\begin{center}
\begin{minipage}[t]{0.48\textwidth}
\begin{alltt}
    module rear_propeller() \{
        for (i = [0:2]) \{
            rotate([0, 0, i * 120])
            translate([0, 8, 60])
            scale([0.3, 1, 0.1])
            sphere(10);
        \}
        cylinder(h=60, d=4);
        translate([0, 0, 60])
        difference () \{
            cylinder(h=1, d=40);
            translate([0, 0, -1])
            cylinder(h=10, d=38);
        \}
    \}
\end{alltt}
\end{minipage}
\hfill
\end{center}
}

Throughout this process, the cross-representation highlighting mechanism allowed Alex to independently create and refine a complex 3-D model. Each module of the helicopter model was also called and arranged using the same iterative process until they were correctly glued together. The final code block for the model assembly is also shown below:

{\small
\begin{verbatim}
    body();  // Main helicopter body
    translate([-12, 0, -20])
    landing_gear();  // Left landing gear
    translate([12, 0, -20])
    landing_gear();  // Right landing gear
    translate([0, 0, 30])
    main_propeller();  // Top propeller positioned above body
    rotate([90, 0, 0])
    translate([0, 0, 60])
    rear_propeller();  // Tail propeller
\end{verbatim}
}

However, despite the success in creating an acceptable helicopter 3-D model, the final artifact also has some problems, indicating that future improvements are needed. The main rotor blade assembly is improperly positioned above the body, failing to connect securely to a defined rotor hub. Similarly, the tail rotor extends too far from the fuselage and appears disconnected from the tail boom structure. These misalignments suggest difficulties in establishing proper spatial relationships between components during the modeling process.

\begin{figure*}[h]
    \centering
    \includegraphics[width=0.8\textwidth]{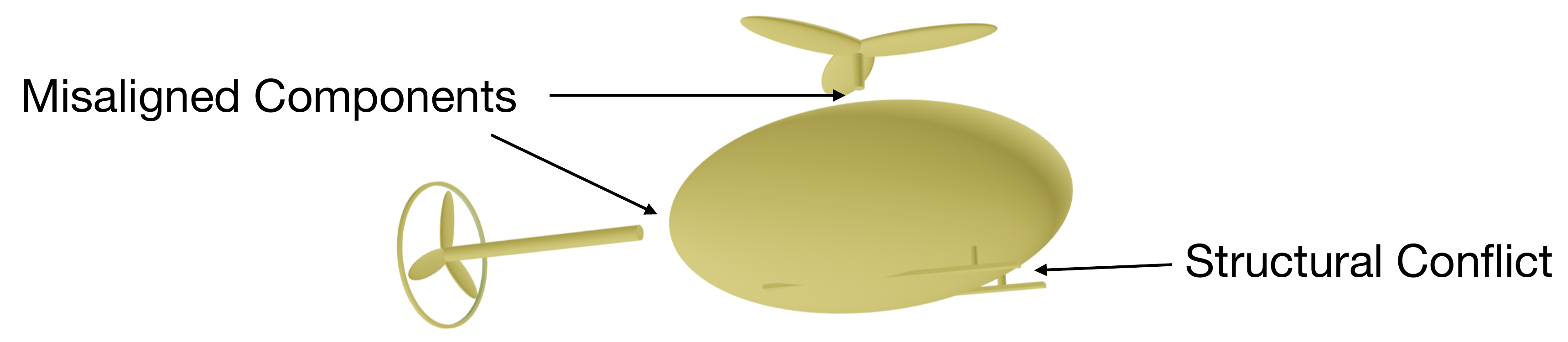}
    \caption{The final artifact of Alex's helicopter model, adopted from our real user studies. It shows misaligned components (main and rear propellers positioned incorrectly) and structural conflict (landing gear intersecting with the body).}
    \label{fig:final_model}
    \Description{A 3-D rendering of a yellow helicopter model with design problems highlighted by arrows and labels. Two issues are identified: "Misaligned Components" pointing to both the main propeller at the top of the helicopter and the rear propeller with ring that extends too far from the body, and "Structural Conflict" pointing to where the landing gear inappropriately intersects with the helicopter's elliptical body. The image illustrates common modeling errors that blind and low vision users might need to identify and correct when using the A11yShape tool.}
\end{figure*}

\subsection{Participatory Design}

We developed A11yShape through close collaboration with a BLV co-author, GK. He is proficient in programming and using screen readers, but he did not have prior 3-D modeling experience. Our participatory design process involved two iterative rounds of user testing. At each session, GK rigorously interacted with system developments, performing representative tasks such as reviewing existing OpenSCAD models, performing code edits with AI assistance, and exploring alternative interface designs. This collaboration yielded detailed usability and accessibility feedback, as well as higher-level insights regarding optimal prompt formulations for effective AI responses, panel navigation strategies, and interface layout improvements. GK recommended adjustments such as more systematic hierarchical navigation across panels, which inspired our cross-representation highlighting mechanism. He also recommended consistent semantic headings optimized for screen reader navigation and interface rearrangements to better accommodate natural workflow sequences during iterative modeling. For example, the code change list was originally put together in one area summarizing all code changes made by BLV users themselves or the AI in a chronological order. After the participatory design, this change list was separated: one list was placed in the code editor showing changes made by humans, and the other list was placed in the AI feedback panel showing changes made by AI, ensuring that users could clearly distinguish between their own modifications and AI-suggested changes, reducing cognitive load when tracking model evolution. This separation also prevented potential confusion between human and AI authorship, supporting clearer attribution and more intentional acceptance of suggested changes. Subsequently, we refined ARIA labels, defined semantic header hierarchies clearly, and restructured panel ordering to realize a logical progression from initial drafting, verification, and final modifications.

%% file: sections/4-study.tex
\section{Study Method}
We conducted a remote, multi-session exploratory user study involving four BLV participants to evaluate the usability of A11yShape and to understand their interactions and evolving strategies during AI-assisted 3-D modeling tasks. Additionally, we investigated workflows and strategies BLV users develop when navigating a multi-representational, spatial modeling environment. In our study, we did not include a baseline comparison with existing 3-D modeling tools like Blender or Fusion 360 because existing tools are effectively unusable with screen readers, making such comparisons uninformative. Instead, we focus on how A11yShape enables first-time independent access to this domain.

\subsection{Participants}

We recruited four BLV participants (all male, ages 21--32, with an average age of 24.3) through prior outreach contacts. The participants varied in vision status, with one fully blind (no light perception), one with minimal perception sufficient for distinguishing day from night, and two various degrees of low vision. All participants regularly relied on screen readers to interact with digital content and had programming experience from one to more than ten years along with familiarity using AI-based conversational agents developed recently. Participants had no prior experience with OpenSCAD or other 3-D modeling tools. We intentionally recruited BLV users with existing programming and AI familiarity to reflect realistic use-case scenarios involving BLV programmers or designers navigating unfamiliar but related technical domains (\textit{i.e.,} computational modeling). Our study was approved by our institutional review board, and we received informed consent from all participants.

\subsection{Apparatus}
Participants interacted via video conferencing and accessed A11yShape on their own desktop computers equipped with their preferred screen readers and code editors. A11yShape provides a web interface compliant with standard screen reader navigational guidelines (WCAG) and integrates OpenSCAD. Each interaction was logged, and video conferencing software recorded audio, screen interactions, and spoken-aloud thought processes with participant consent.

\subsection{Procedure}
Our study included three separate sessions for each participant (2.5 hours per session, totaling approximately 7.5 hours per participant). We intentionally spaced sessions several days apart over an 11-day period, allowing participants time to reflect, adapt, and progressively build familiarity with both OpenSCAD coding syntax and the A11yShape features. Each participant received compensation of \$50 per session. 

\subsubsection{Session 1: Tutorial and Introductory Tasks}
Participants first learned basic OpenSCAD syntax and were introduced to A11yShape features, including model descriptions, hierarchical component navigation, AI-assisted verification loops, and version control. They explored A11yShape using standardized introductory modeling tasks, including creating simple geometric shapes and examining a pre-prepared, complex 3-D bacteriophage model. Given the introductory nature of this session and the complexity of this bacteriophage model, participants were only asked to ``read'' this model through A11yShape's features instead of further editing it for other purposes. Participants interacted freely with the system and reported usability impressions through semi-structured interviews. Based on participant feedback, we iteratively refined the interface and system functionalities prior to the next session.

\subsubsection{Session 2: Guided 3-D Modeling Tasks}
After reviewing system updates and briefly revisiting OpenSCAD fundamentals, participants independently completed two guided 3-D modeling tasks of increasing complexity. Participants selected from predefined prompts (a Tanghulu\footnote{A sugar-coated hawthorn skewer, famously known as a traditional Chinese desert} model with a bite mark for the first task; either a standing robot or a robot-trailer assembly for the second task). To isolate system usability from programming challenges, experimenters provided OpenSCAD syntax assistance upon request but offered no hints regarding system operation.
Each task provided detailed natural-language descriptions of the models without visual references, enabling participants to comprehend requirements without visual cues. Tasks were allotted one hour each, and participants were encouraged to verbalize their reasoning processes throughout. The first task focused on basic modeling capabilities, while the second task emphasized more complex spatial relationships between multiple components. After each task, we conducted brief semi-structured interviews to gather immediate feedback on the participants' experiences and challenges encountered.

\subsubsection{Session 3: Free-form Creative Modeling}
In the final session, we asked participants to independently select and build unique 3-D modeling projects reflecting their own interests or creative ideas, closely simulating real-world 3-D modeling scenarios. Similarly, participants could freely query the AI assistant provided by A11yShape and request support only in OpenSCAD coding syntax. We explicitly aimed to observe unstructured exploratory interaction patterns, creative modeling strategies, and the perceived value of the system for open-ended design tasks. Participants described their intended goal before modeling and communicated their thoughts during interaction. Post-task interviews again collected participants' qualitative feedback on the entire modeling process and their experiences throughout the study. We also gathered their responses on system usability scale \cite{bangor2008empirical} scores. 

\subsection{Analysis}
We transcribed all session recordings and conducted thematic analysis \cite{clarke2017thematic} on the qualitative data, including think-aloud transcripts, user interviews, interaction behaviors, and observational notes. Two researchers independently performed open coding, line-by-line, to generate initial codes capturing interactions with system features, workflows, strategies, and demonstration of creative autonomy during modeling. These initial codes were collaboratively discussed and refined, leading to the iterative development and finalization of a comprehensive, hierarchical codebook. The codebook themes include participants' reactions to the overall system, code editor, responses provided by AI, version control, hierarchical representation of the model, and cross-representation highlighting mechanism; participants' performance on tasks, their challenges, as well as suggestions to the system. Using the finalized codebook, the same two researchers independently re-coded all transcripts and notes. Intercoder disagreements were thoroughly discussed and resolved collaboratively to ensure consistency. Additionally, we quantitatively analyzed interaction logs that captured participants' usage of specific A11yShape functions (hierarchy navigation, verification loops, version control) to further triangulate our qualitative findings and reveal evolving user behavior patterns over time.

%% file: sections/5-result.tex
\section{Results}

We report key insights on BLV participants' interactions and experiences with A11yShape. 
We first report participants' performance and showcase created artifacts. Then we discuss participants' impressions and perceived effectiveness of the system, followed by identified challenges, findings of how users engaged with A11yShape, including their workflows and distinctive user-developed strategies.

\subsection{Performance and Artifacts}
\begin{figure*}[h]
    \centering
    \includegraphics[width=\textwidth]{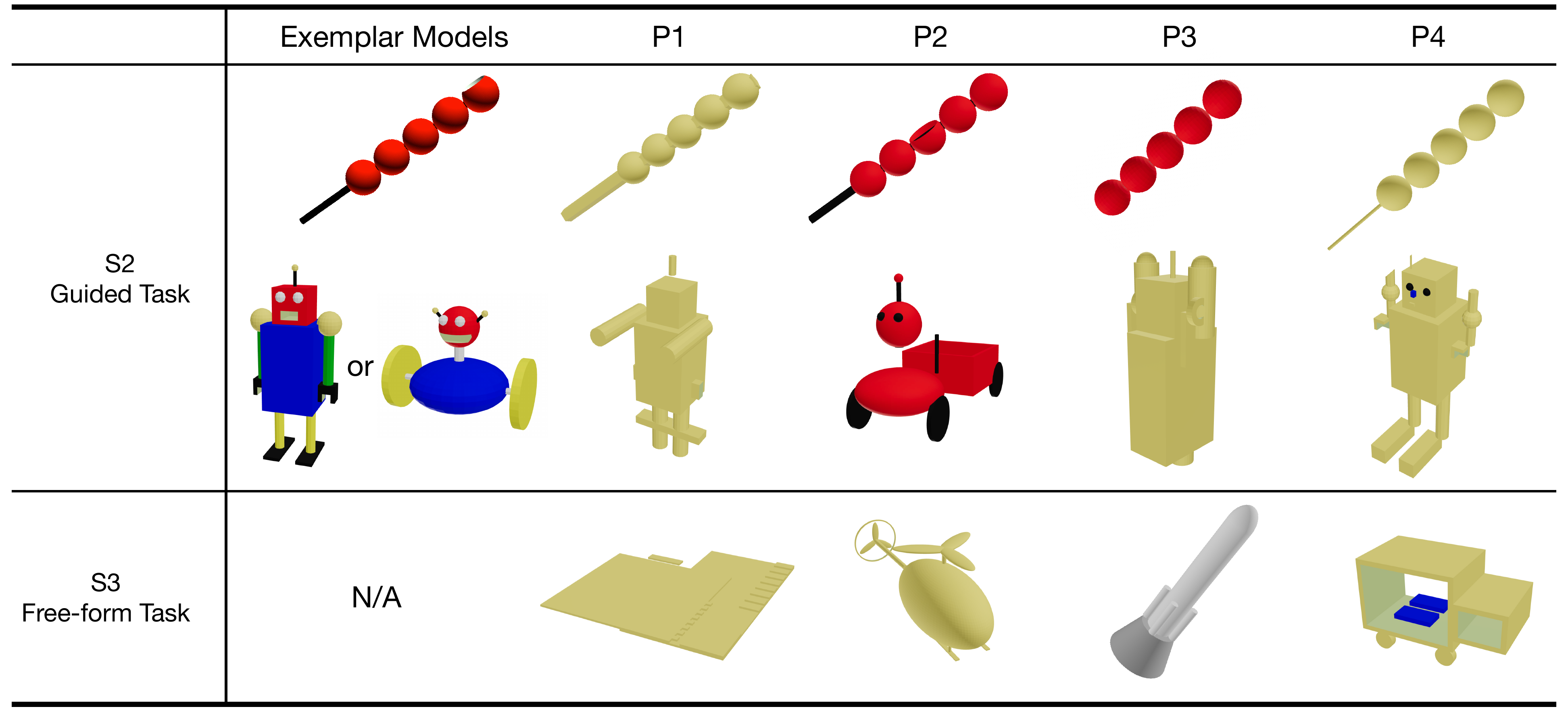}
    \caption{3-D modeling outputs across two task conditions: S2 (Guided Task) showing exemplar models and participant creations of simple objects with primitive shapes, and S3 (Free-form Task) displaying participant-designed complex models including a circuit board, a helicopter, a rocket, and a cart.}
    \label{fig:artifacts}
    \Description{A table comparing 3-D modeling outputs across four participants (P1-P4) in two task conditions. The top row shows S2 (Guided Task) results where participants created models based on exemplars shown in the leftmost column--- simple objects like skewered spheres, robots, and toy vehicles. Each participant's interpretation varies in color (red or yellow) and detail. The bottom row shows S3 (Free-form Task) results where participants created original designs: P1 made a circuit board, P2 created a helicopter with propeller, P3 designed a rocket, and P4 built a cart with a blue component. The first exemplar column shows "N/A" for the free-form task since no reference models were provided. This visualization demonstrates how blind and low vision users progress from guided to creative 3-D modeling with A11yShape.}
\end{figure*}

All four participants successfully completed both guided and free-form 3-D modeling tasks using A11yShape (Figure~\ref{fig:artifacts}). Across the sessions, participants independently created 12 distinct models, demonstrating the system's capacity to support both structured and open-ended modeling workflows.

In the guided task (Session 2), all participants modeled a Tanghulu skewer, though with slight individual variations. P1 and P4 did not apply color to the hawthorn balls, and P3's skewer was shorter than intended. Only P2 implemented the optional ``bitten'' effect by subtracting a half-sphere using OpenSCAD's boolean \codeword{difference()} function, indicating comfort with more advanced operations. For the second guided task, participants were asked to construct a robot. P1, P3, and P4 created a standing robot; P2 extended the concept to a robot-trailer assembly. P2 and P4 added facial features, enhancing the expressiveness of their models. Except P2 who added his own preferred colors to the robot model, other participants stopped at creating the models and did not color the model components given time constraints.

In the free-form task (Session 3), participants exercised creative freedom to design their own models. P1 created a circuit board with surface details. P2 constructed a helicopter, discussed in detail in the user journey section. P3 created a rocket with added symbolic elements (\textit{e.g.,} national flag and rocket name), although these were omitted in the final exported artifact. P4 designed a wheeled cart with distinguishable components.

Despite overall successful creation of artifacts, we observed structural and alignment challenges in participants' models, particularly in more complex designs. For example, in P2's helicopter (which was used in the user journey in Figure~\ref{fig:final_model}), several components showed misalignment issues where the propellers were not perfectly connected to or aligned with the main body. Additionally, the model revealed structural conflicts where supporting elements intersected incorrectly with the fuselage. These issues reflect common challenges in non-visual 3-D modeling, where spatial relationships between components must be managed without visual feedback. In the free-form modeling tasks, participants typically considered models complete once they generally matched their conceptual expectations, leaving these more nuanced alignment and structural issues unaddressed.

Participants reported a mean System Usability Scale (SUS) score of 80.6, suggesting high perceived usability. Notably, P2 clarified that his relatively lower score stemmed from unfamiliarity with OpenSCAD syntax rather than limitations of the system itself, emphasizing that he found A11yShape more helpful than the numerical rating alone suggests. 

\subsection{Impressions and Experience}
All four participants noted that the system fundamentally shifted their perceptions about their ability to create and manipulate 3-D objects, as they previously thought that such tasks were impossible for blind users and users with no prior experience:

\begin{quote}
    \textit{I had never modeled before and never thought I could. Due to various issues, I didn't have an overall understanding of models and didn't know how to construct them. However, through today's simple modeling process, it provided us (the BLV community) with a new perspective on 3-D modeling, demonstrating that we can indeed create relatively simple structures. (P4)}
\end{quote}

Similarly, P2 described the experience as ``revolutionary.'' Particularly, participants highlighted that A11yShape provided an intuitive interface, including helpful AI-generated descriptions, version control for managing changes, and the model parts hierarchy for pinpointing model elements. They expressed satisfaction with the system's ability to realize their design concepts, especially the potential in supporting spatial understanding and automating tedious tasks like syntax and grammar checks. Participants also positively acknowledged core components, especially highlighting the accessibility of the code editor and the AI-generated descriptions that compensated for the lack of visual verification. P2 stated, ``\textit{The editor is one of the most satisfying parts of the system... very accessible to use.}'' Furthermore, they also recognized A11yShape as innovative and beneficial beyond immediate use, highlighting a broader implications for BLV education:

\begin{quote}
    \textit{A11yShape could significantly help visually impaired children build a better understanding of the physical world. (P3)}
\end{quote}

Participants also recognized the value of the cross-representation highlighting mechanism specifically, which enabled them to fluidly shift between different ways of understanding the model. As P4 described, ``\textit{I already had an understanding of the model's composition from the AI description, but used the Hierarchical List to better understand specific structures.}'' Additionally, the ability to locate code associated with specific components streamlined the editing process. P4 noted, ``\textit{After making changes, if I want to make further adjustments, I can directly find where I made the previous changes and press Enter to locate it.}'' These cross-linked representations supported BLV user's autonomy and minimized the friction of navigating between multiple non-visual representations (\textit{e.g.,} code, descriptions, and semantic hierarchy) that sighted users typically access instantly through visual feedback.

Despite these positive experiences, participants expressed dissatisfaction around several key areas. First, heavy reliance on textual descriptions created cognitive overload in longitudinal modeling tasks. For instance, P4 critically noted receiving excessive or redundant information from the AI, making it ``hard to locate key details,'' which significantly slowed the iterative modeling processes.
Participants also indicated that OpenSCAD's specialized syntax posed initial difficulties, especially for newcomers to programming-based modeling.
Participants also identified inherent limitations in relying on textual interactions without tactile feedback. When examining spatially complex structures like the bacteriophage model, P3 expressed frustration: \textit{``Without touching it, no matter how detailed the description, there's essentially no way to genuinely visualize the entire model.''} He described this limitation as potentially ``unsolvable'' without complementary tactile support.
In addition, several participants raised concerns related to uncertainty and occasional inaccuracies in AI responses, particularly regarding quantitative details such as precise element counts. This was notably frustrating for P4, who experienced that AI generating different counts of objects for a same model, undermining trust in accuracy.

\subsection{Challenges}
\label{sec:challenges}
Beyond their overall experiences, we also observed and listed detailed challenges of how blind users are engaging with this highly interpretive and visual-spatial workflow. Our findings in participants' challenges align well with prior work which explored blind users' experience with other visual artifacts like artboards \cite{schaadhardt2021understanding}.

\subsubsection{High Cognitive Load due to Dynamic Representations}
Participants experienced substantial cognitive demands associated with A11yShape's dynamic, four-layer representation of 3-D models (the rendered 3-D model, OpenSCAD code, AI description, and the semantic hierarchy). Due to the inherent linear interactions with the system using screen readers, BLV participants must sequentially process textual representations, which can contain redundant information not required each time. For example, P4 noted instances involving unnecessary textual details: ``\textit{I sometimes only need code-related descriptions, yet still receive irrelevant model descriptions.}''

The iterative modeling process further compounded this challenge, as participants needed to recall and differentiate among textual histories representing multiple model states. Although the version control feature partially mitigated this challenge, participants indicated difficulties retaining mental models across iterative cycles.

\subsubsection{Difficulty Understanding Spatial Relationships}
The accurate comprehension of spatial structures and components was consistently problematic for participants. Challenges arose primarily from difficulty estimating relative positions, proportions, and coordinates. P2 highlighted this explicitly: ``\textit{The main problem I'm encountering is (mentally) calculating spatial proportions and coordinates},'' while another described challenges adjusting complex interdependent structures, such as the bacteriophage model.
Participants highlighted the necessity of precise mental calculation when translating and rotating model components, which proved especially challenging when designing interconnected or composite structures. Despite overcoming many spatial challenges through practice, maintaining spatial coherence while modifying individual components remained difficult throughout.

\subsubsection{Constructing \& Maintaining Mental Models}
The absence of direct visual or tactile feedback significantly affected participants' ability to develop working mental models throughout iterative editing. Participants indirectly confirmed correctness through AI responses rather than direct verification through visual inspection. P3 explained: ``\textit{I judge based on the system's descriptions as indirect verification... assuming minimal deviation if the response aligns roughly with my understanding.}'' P3 also noted when he tried to create a spiral model: ``\textit{I could understand the presence of a spiral structure from the AI's description and modify it accordingly, but I found it difficult to mentally visualize the shape of the spiral.}'' 

\subsubsection{Uncertainty of Operation Success}
Participants frequently expressed uncertainty regarding the success of their modeling operations, influenced by cognitive challenges and system limitations. Without visual or direct tactile feedback, participants often voiced strong hesitations about independently managing intricate spatial decisions. For instance, when manually adjusting component dimensions, P2 explained, ``\textit{If I were to manually adjust the height, I probably wouldn't be able to calculate it accurately...}'' This uncertainty was not limited to spatial calculations; participants also expressed psychological hesitation in refining or adjusting existing model structures. For example, P3 articulated significant anxiety around further adjustments, noting, ``\textit{I feel there are definitely some issues... One major reason preventing me from refining the structure is the fear that AI modifications might damage the existing structure. I think this is a psychological issue.}'' Such concerns reveal an inherent barrier to fully confident interactions with 3-D modeling, highlighting a critical challenge that BLV users face when exploring adjustments and refinements without direct visual validation.

\subsection{Workflows}

With different levels of AI usage, participants developed different workflows when they create 3-D models with A11yShape.

We identified three workflows involving different degrees of AI usage and user intervention. The first workflow involved predominantly independent coding, with users employing the AI primarily to obtain descriptive feedback, perform verification, and assist with minor modifications. For instance, P4 adopted this approach due to concern that AI-generated models might deviate significantly from intended design objectives:
\begin{quote}
``\textit{AI might not fully grasp my intentions. Relying on AI to create the primary model might not align with my vision.}''
\end{quote}
While providing users greater control and fostering a clear understanding of the entire design process, we observed this workflow required more substantial manual cognitive effort and occasionally slowed progress, particularly when tackling unfamiliar or complex spatial structures.

The second workflow, most common among participants, balanced AI support and manual refinement. Participants initially leveraged the AI system to produce the overall model framework quickly, subsequently using direct code edits to refine and adjust finer details. Participants favored this workflow for effectively balancing efficiency against adequate creative control and accuracy. P2 highlighted its practical advantage:
``\textit{This approach generally worked well and saved a significant amount of time.}''
However, manual involvement remained necessary due to occasional limitations or inaccuracies in AI-produced outputs, underscoring the necessity of human input in complex, detailed aspects.

In contrast, the third workflow involved complete reliance on AI for model generation. For example, P3 entirely delegated modeling to the AI, minimizing manual coding intervention. Although this approach notably increased efficiency and speed, it introduced critical drawbacks, such as diminished familiarity with AI-generated structures and reduced capability to intervene when the AI reached performance limits. Recognizing these limitations, P3 subsequently considered transitioning towards a more balanced approach that included manual interaction.

\subsection{Strategies and System Usage}
We also report notable findings on how users interacted with A11yShape's features and developed their own workaround to use the system. 

\subsubsection{Incremental Building Through AI-verification Loop}
Participants commonly built models incrementally using repeated cycles of AI-driven verification involving model interpretation and code refinements. This approach allowed progressive validation and reduced cognitive complexity by limiting the scope of change within each step. P2 described this strategy through analogy: ``\textit{Modular modeling is like building blocks. It separated considerations of spatial relationships, initially focusing only on dimensions and shapes, with spatial relationships addressed during assembly.}'' This strategy is also rooted from the inherent challenge of how participants felt uncertain towards changing model parts actively and the high cognitive load of performing holistic changes over the entire model (Sec. ~\ref{sec:challenges}).

As P4 described, ``\textit{After each modeling session, I wait for AI's description because it's actually quite important.}'' This reflects the central role of AI as a verifier in the absence of visual feedback. Similarly, P4 noted that iterative feedback helped him detect design inconsistencies: ``\textit{The AI helps me mentally construct models I was previously unsure about or didn't know how to modify... and detect differences between successive models.}'' 

\subsubsection{Semantic Hierarchy and Version Control for Error Correction}
Participants actively leveraged semantic hierarchical navigation and version controls to identify and correct errors. This allowed users like P3 to effectively ``\textit{backtrack through previous states}'' if AI-based or manual modifications did not align with their expectations, enabling systematic correction without losing orientation within the more complex structure.

Version control served as a critical safeguard during exploratory modeling. As P3 reflected, ``\textit{I shouldn't have used undo, I should have backtracked through the version history.}'' This underscores the need for structured correction tools that offer better control than traditional linear undo operations. P4 similarly appreciated the ability to review changes: ``\textit{It was helpful for reviewing what I had done after modifying the model.}''

Semantic hierarchy also played a vital role in allowing targeted navigation and served as a ``middle-layer'' between the accurate but abstract codes and the rendered but inaccessible 3-D model. P4 noted, ``\textit{The hierarchical list made the model's structure clearer and more targeted},'' while P2 appreciated its recursive structure: ``\textit{It helps me grasp the model's composition by showing levels of structure.}'' These features enabled participants to locate and revise specific parts of a model without having to sequentially process the entire codebase, which is an important accessibility affordance for screen reader users.

\subsubsection{Real-world Metaphors for Mental Model}
Real-world metaphors from AI descriptions became integral to participant understanding during mental model construction. Users explicitly valued AI-generated metaphors, provided they corresponded to objects familiar through prior tactile experience. As P4 noted, ``\textit{One good aspect is that sometimes using real-life examples makes it easier for me to understand.}'' These metaphors served as anchors that helped translate abstract shapes into mental representations grounded in everyday experiences.

However, mismatched metaphors could constrain or even distort mental model construction. P3 shared a notable instance of misalignment, saying, ``\textit{I found the analogy between the bacteriophage and a spider confusing, as I had never touched a spider and couldn't picture its shape.}'' This underscores the importance of grounding metaphors in objects that users have physically encountered, especially within the BLV community where tactile experience plays a dominant role in spatial understanding.

Participants also emphasized the balance between metaphorical clarity and precision. When metaphors aligned well with prior tactile knowledge, they became powerful tools for both comprehension and confidence-building. Conversely, unfamiliar analogies introduced ambiguity, potentially leading to misconceptions about a model's structure or form.

\subsubsection{Trust-building Process with the AI Assistant}
Participants incrementally developed trust in the AI assistant through repeated experience of consistent alignment between expected and actual AI outputs. Initially, participants exhibited caution and limited their reliance on AI-generated content due to uncertainty about its accuracy and reliability. In particular, they experienced unsatisfactory outcomes when requesting complex OpenSCAD models in a single interaction. We attribute these difficulties partly to unclear participant instructions and partly to inherent limitations in current LLMs, given that OpenSCAD has limited representation in mainstream training corpora. Over time, participants learned to adjust the complexity and granularity of their requests, shifting from broad, single interaction tasks toward incremental requests at the component level. By consistently verifying AI-generated outputs at this smaller scale, participants gradually accumulated positive experiences. This incremental approach effectively reduced uncertainty about AI reliability and increased user confidence, fostering a more collaborative interaction style between participants and the AI assistant. 

For example, during the helicopter free-form task, P2 initially attempted to generate the entire helicopter model by providing a detailed, paragraph-length description of all components at once. This approach was unsuccessful: the generated output did not align with his intent, and the AI provided a description unrelated to a helicopter. Consequently, P2 became skeptical of the AI's capability to handle complex modeling instructions in one step. Following this experience, P2 adjusted his strategy by decomposing the helicopter modeling task into smaller components. He first instructed the AI to generate a simple sphere as the helicopter's fuselage, gradually adding other elements in subsequent requests, such as the tail, rotor, landing gear, and propeller. In adopting this modular modeling approach, P2 consistently verified and confirmed correctness for each new element before proceeding to the next. This iterative cycle allowed him to clearly identify and quickly rectify any discrepancies, reducing uncertainty regarding AI reliability. By incrementally achieving successful outcomes, P2 rebuilt confidence in the AI assistant and developed a more effective collaborative interaction style.

Overall, trust formation reflected an iterative ``calibration process,'' in which participants continuously compared the assistant's outputs against personal expectations and mental-model accuracy. Successful verification cycles reinforced confidence, whereas deviations (\textit{i.e.,} misunderstandings or inaccuracies) damaged trust and prompted more cautious subsequent interactions. Participants ultimately arrived at productive, balanced collaboration models by actively discovering and navigating AI capability boundaries: strategically decomposing challenging tasks, verifying model outputs rigorously, and maintaining appropriate critical independence to accommodate AI uncertainty.

%% file: sections/6-discussion.tex
\section{Discussion}
In this section, we zoom out to offer a summary of results and feedback about A11yShape from our user studies. We also discuss A11yShape's limitations and avenues for future work in this research effort.

\subsection{Summary of Results}
Our study showed that BLV participants were able to independently create both structured and free-form 3-D models using A11yShape, challenging common assumptions about the inaccessibility of spatial design. Participants highlighted the system's cross-linked representations and AI feedback as especially empowering, enabling them to understand, build, and revise models without vision or with low-vision. While participants faced challenges with spatial reasoning and cognitive load, they developed creative strategies like modular modeling and iterative AI verification to navigate them. Across sessions, users increasingly trusted the system and shifted from cautious exploration to confident modeling. 

\subsection{Cross-Representation Highlights}
We draw attention to the historical roots, design rationale, and broader potential of A11yShape's core interaction feature: the \textit{cross-representation highlighting} mechanism. This design emerges from long-standing accessibility research and assistive technology development \cite{schaadhardt2021understanding, designchecker, pengchi2023slidegestalt, zhang23a11yboard, liu2022crossa11y, chang24editscribe}, aiming to address a fundamental challenge for BLV users---the mismatch between how information is visually structured and how it is linearly conveyed through screen readers.

This gap is especially evident in tasks involving rich, multimodal media. BLV users interact with content types ranging from static images and formatted documents to dynamic artifacts like charts, slides, videos, websites, and, in this work, 3-D models. Some of these formats follow WYSIWYG (what-you-see-is-what-you-get) conventions, while others, like data visualizations or websites, can be partially accessed through their underlying textual structures. However, complex artifacts such as videos or 3-D models often require additional layers of representation, like semantic hierarchies or generated descriptions and captions, to be navigated non-visually. 

A11yShape addresses this longstanding disconnect through the proposed cross-representation highlighting: a mechanism that synchronizes focus and interaction across multiple parallel representations of a component---code, rendering, semantic tree, and more. This dynamic linkage enables rapid multimodal navigation and reduces the cognitive overhead of switching contexts. For example, selecting a component in the hierarchy view highlights the corresponding code block and rendered geometry, offering BLV users consistent spatial or semantic anchors across modalities.

Although similar ideas have appeared in prior work, they have rarely been treated as a unified interaction pattern. For example, DesignChecker \cite{designchecker} and EditScribe \cite{chang24editscribe} similarly explore alternative representations to support navigation and editing of inherently visual content like websites and images. We argue that cross-representation highlighting is a generalizable accessibility technique with wide applicability across creative domains. From authoring slide presentations to editing websites or designing rich data visualizations, this approach can enhance nonvisual access by tightly coupling descriptive, structural, and visual representations. A11yShape contributes a concrete instantiation of this mechanism in 3-D modeling, while also pointing to its broader potential in accessible creative workflows.

\subsection{Feedback and Suggestions for Improvement}
Participants provided valuable feedback for enhancing system capabilities throughout our study. They requested more refined AI descriptions that would better communicate spatial relationships, prioritize critical details in concise formatting, and consistently connect designs to tactilely familiar real-world objects to facilitate effective mental mapping. This emphasis on tangibility extended to their strong desire for complementary tactile interfaces that would enable direct mental visualization and verification, addressing the inherent abstractness of purely language-based descriptions.

For the code editing experience, participants suggested implementing optional auto-completion features and more accessible ways to access standard functions, potentially through low-code or no-code components that would reduce the technical barriers to model creation. They also recommended repositioning the hierarchical views within the interface layout to facilitate more frequent and seamless consultation, alongside implementing collapsible formats that would improve readability during extended modeling sessions.

Participants with residual vision requested additional visual confirmatory features, such as multiple perspective renders or reference lines that would aid approximate spatial understanding and verification. They emphasized how having different viewing angles could significantly enhance their ability to comprehend and validate complex spatial relationships within their models.

Looking toward future applications, participants expressed considerable interest in practical integration scenarios, particularly incorporating A11yShape within educational resources targeting visually impaired children. They envisioned leveraging the system to improve spatial cognition skills through digital modeling, with the added benefit of producing tangible outcomes through 3-D printing. 

\subsection{Limitations}
Despite its promise, A11yShape is constrained by several limitations. First, the system's AI assistance is bounded by the capabilities of current LLMs. While tools like GPT-4o provide general-purpose reasoning, they often struggle with writing accurate and efficient OpenSCAD code, which is likely due to the niche nature of the language and its limited presence in mainstream training corpora. 

Second, the modeling experience in A11yShape is most effective when models are constructed from simple and interpretable geometric primitives. As model complexity increases, particularly for designs that cannot be easily broken down into modular or semantically meaningful units, users face increasing difficulty. Current workflows do not yet support the kinds of abstraction or decomposition required for more intricate models.

Third, the system lacks proactive detection and correction mechanisms for problems like structural and alignment issues in 3-D models. Without a built-in design validation for sighted users, A11yShape currently operates as a passive tool rather than an active agent that identifies potential problems. As observed in participants' artifacts (\textit{e.g.,} P2's helicopter), misaligned components and structural conflicts often remain undetected until the final stages of modeling, if at all. Without automated spatial relationship validation, BLV users must rely solely on their mental models to track component positioning and intersections. 

Additionally, our study was limited in scope to a single guided task and a single free-form modeling session per participant. While these sessions were much longer than standard usability testing sessions and provided valuable insights into A11yShape's perceived value and supported workflows, they do not reflect sustained or longitudinal use. As a result, we were unable to capture how users' strategies, preferences, and creative practices might evolve over time, nor could we assess long-term learning effects or potential fatigue. We also did not compare BLV users' performances with existing technologies (\textit{e.g.,} just code editor and a LLM for chatting) in a quantitative way to further showcase the usability of A11yShape. Future work should explore extended deployments to better understand how A11yShape integrates into users' broader modeling habits and whether the system supports the development of more advanced modeling skills over repeated use.

Finally, and more fundamentally, the absence of tactile feedback remains a core limitation for BLV users. While A11yShape significantly streamlines the modeling process, it cannot fully replace the physical interaction that comes with holding a printed 3-D object. Theoretically, an iterative loop where users create a model (manually or through AI), print it, and refine it based on touch feedback \cite{shi_chi19, shi2017markit, siu2018shapeshift, touchpilot} could make 3-D modeling accessible. However, this process is time-intensive and financially burdensome, especially for independent or novice users. Our system reduces the need for repeated printing, but the gap between abstract spatial reasoning and tactile experience remains.

\subsection{Future Work}
Future work should explore ways to bridge the gap between virtual modeling and tactile confirmation. One promising direction involves integrating A11yShape with physical prototyping workflows. For example, a final confirmation step could involve 3-D printing the model, paired with computer vision techniques that detect user touch and interaction on the printed object. These methods, which are demonstrated in prior work on interactive 3-D models \cite{shi_assets17, shi_chi19, i3ms, leporini_taccess20}, can offer real-time, spatially grounded feedback to users as they explore the physical model. 

Another potential future work avenue is developing proactive model validation systems that can detect structural issues and misalignments. It would transform A11yShape from a passive editing tool into an intelligent assistant that actively guides BLV users toward creating not only usable but also structurally sound models.

In addition, improving the AI assistant's modeling capability remains an open challenge. Training or fine-tuning LLMs on domain-specific corpora, such as OpenSCAD examples or accessible modeling tutorials, may increase their accuracy and reliability. Coupled with low-code or no-code extensions suggested by participants, future iterations of the system could better support novice users while reducing the need for precise syntax.

Lastly, extending A11yShape beyond individual use cases into educational and collaborative contexts presents valuable opportunities. By supporting shared modeling tasks and enabling educators to guide visually impaired students and programmers through structured modeling exercises, the system could play a broader role in improving spatial cognition and design literacy for BLV learners.

%% file: sections/7-conclusion.tex
\section{Conclusion}

We have presented A11yShape, an accessible 3-D modeling system that enables BLV users to create, edit, and verify complex models through synchronized representations of code, structure, and AI-generated feedback. Through participatory design and a multi-session user study, we found that BLV programmers could meaningfully engage in 3-D modeling tasks, develop individualized strategies, and build confidence in navigating visual-spatial workflows. A11yShape presents an initial step toward making 3-D modeling---already a cognitively demanding task for many sighted users---accessible to BLV users. At its core, the system addresses a fundamental challenge also observed in prior work: the gap between how screen reader users perceive information linearly (1-D) and how 3-D models are rendered and represented across multiple spatial and semantic formats. By introducing dynamic cross-representation highlighting and multimodal verification loops, A11yShape helps bridge this perceptual divide. While limitations in LLM capabilities, spatial abstraction, and the absence of tactile feedback remain, this work lays the foundation for future accessible modeling systems that combine virtual modeling with physical interaction, more robust AI instruction, and deeper spatial reasoning support.

\section{Contribution Statements}

Zhuohao led the paper writing and presentation, with contributions from Liang, Jacob, Anhong, Mingming, Angus, Haichang, and Chun Meng. Research ideation was led by Liang, with input from Zhuohao, Anhong, Faraz, and Angus. The A11yShape system design was led by Haichang and Chun Meng, who proposed and iterated on key functional features aligned with the research vision; Zhuohao and Liang provided iterative feedback, and Faraz and Anhong contributed. Haichang and Chun Meng led the system development, with support from Gene. Gene also participated in the participatory study as a BLV user, providing critical feedback; Chun Meng, Haichang, Zhuohao, and Liang contributed to the participatory study. The validation study was led by Zhuohao, with contributions from Liang and Mingming. Chun Meng and Haichang led the design of the multi-session study, with input from Zhuohao, Anhong, and Liang. Participant recruitment was supported by Junan and Mingming. Haichang led the planning, coordination, and facilitation of the study sessions, with Chun Meng assisting in session facilitation, Junan co-planning and coordinating, and Anhong and Liang participating in the first session. Zhuohao led the data analysis, with Haichang and Chun Meng proposing the initial codebook. Junan supported time-series data collection, and Haichang and Chun Meng assisted in compiling evidence for initial analysis. Haichang produced the video.

\section{Acknowledgment}
We would like to thank Dr. Venkatesh Potluri for sharing his first-hand experience with 3D modeling, which inspired the idea of this work in 2021. Zhuohao (Jerry) Zhang was supported by the Apple Scholars in AI/ML PhD fellowship.

%% file: sections/99-appendix.tex
\appendix

\section{Prompts}

\subsection{Describing 3-D Models}
\label{sec:prompts_describing}
We asked LLMs to describe 3-D models from different perspectives, including from the codes, from rendered images, specific components, comparison between model versions, etc.\\

\noindent\textbf{Describing Models from Images}

\begin{lstlisting}
As a good 3D model descriptor, you will receive images from the OpenSCAD 3D model and generate a detailed description of the 3D model, describing what the 3D model is and what parts it consists of. After that, you will work with the code interpreter to match the different parts of the model to the code that generates this corresponding part.
Use the following format for output:
***Report Begins***
##Description of the model##
[Insert the description of the model here, highlighting key elements.]

##Summary of the model##
[Insert the summary of the model here, contains all the components.]
***Report Ends***
\end{lstlisting}

\noindent\textbf{Analyzing Code for BLV Users}

\begin{lstlisting}
As a code interpreter, you will receive a set of OpenSCAD code and analyze the code for a blind user to understand.
Given the Openscad code, you will analyze the code and provide a detailed description of the code, highlighting the key elements and code structure. After that, you will evaluate the code, highlighting the strengths and weaknesses. 
***Code Begins***
'''openscad'''
{code}
'''openscad'''
***Code Ends***

Use the follow format for output:
***Report Begins***

##Description of the openscad code##
[Insert the description of the openscad here, highlighting key elements and code structure.]

##Summary of the code##
[Insert the summary of the code here, contains all the components.]

##Evaluation of the code##
[Insert the evaluation of the code here, highlighting the strengths and weaknesses.]

##Codes##
"Code1", [Function of the code],[Suggestions for improvement] 
[content of Code1]
"Code2", [Function of the code],[Suggestions for improvement]
[content of Code1]
...
***Report Ends***
\end{lstlisting}

\noindent\textbf{Describing a Model Component}
\begin{lstlisting}
Given the part of a 3D model and its OpenSCAD code, compare this part of the model in relation to the full model such that a blind user could understand it (eg. spatial position, distance, intersection, size, angle, orientation, side in relation to other parts of the model). Describe how this part affects the model's shape. Only if applicable, mention what operation the part is used in and if it's invisible
\end{lstlisting}

\noindent\textbf{Comparing Model Versions}
\begin{lstlisting}
Given the two versions of a 3D model and its OpenSCAD code, with the last {n} images and code referred to as the current model and the first {n} images and code referred to as the previous model, describe the changes between the two versions, focusing on the visual details such that a blind user could understand it (eg. shape, position, posture, pictures).
\end{lstlisting}

\noindent\textbf{General Guidance for Model Descriptions}
\begin{lstlisting}
Given the 3D model and its OpenSCAD code, describe the visual details such that a blind user could understand it (eg. shape, position, posture, pictures).

You must give a one sentence answer or summary first, followed by more details such that a blind user could understand it. The output should not have formatting since it will be read by a screenreader. Do not mention blind users. The images are of the same model at different angles. Do not mention that there are multiple images. Do not describe each angle separately. The description should be based on the images of the model rather than the code.
\end{lstlisting}

\noindent\textbf{Summarizing Full Descriptions}
\begin{lstlisting}
As a summarizer, here is a paragraph for the blind person to read, but it will take a lot of time for the screen reader to read this paragraph. Please use a simple sentence to restate the main points of the speech so that the blind person can get the most important information in a short time.
{text}
\end{lstlisting}

\subsection{Code Interpreter}
\label{sec:prompts_code}
We used a series of different prompts for matching codes with corresponding model components, formalizing code changes, creating a model or improving codes based on user input.\\

\noindent\textbf{Matching Codes with Models}
\begin{lstlisting}
As an expert in OpenSCAD code interpretation, you will receive a set of OpenSCAD code. For a given piece of code, you will work with the 3D model descriptor to connect the different parts of the 3d model and their corresponding code.
Use the following format for output:
***Report Begins***
##Codes##
"Code1", [The corresponding part in the model], 
[content of Code1]
"Code2", [The corresponding part in the model], 
[content of Code1]
...
***Report Ends***
\end{lstlisting}

\noindent\textbf{Tracking Code Changes}
\begin{lstlisting}
Given the previous OpenSCAD code followed by the current OpenSCAD code, output the list of chunks of code that were added, deleted or changed in the format [{"startLine": <the first line number of the chunk in the current code, or -1>, "endLine": <the last line number of the chunk>, "description": <description of what changed>}]. Output only JSON and nothing else.
\end{lstlisting}

\noindent\textbf{Creating 3-D Model}
\begin{lstlisting}
You are an OpenSCAD expert specializing in accessible code generation for individuals who are blind or visually impaired.  Your primary goal is to translate user descriptions of 3D models into functional, efficient, and accessible OpenSCAD code within a single interaction.

Core Responsibilities:

1.  Comprehensive Requirement Analysis:
- Actively listen to the user's description of their desired 3D model.
- Focus on understanding their vision, including the model's overall shape, dimensions, features, and any specific functional requirements.
- If necessary, politely request clarifying details or suggest alternative approaches to ensure a clear understanding of the project scope.

2.  Accessible Code Generation:
- Transform the user's description into precise, well-structured OpenSCAD code that adheres to industry best practices.
- Prioritize accessibility by:
- Employing clear, descriptive variable names and comments.
- Implementing consistent indentation and formatting for seamless navigation with screen readers.
- Utilizing modules and functions to enhance code organization and reusability.

3.  Proactive Guidance and Optimization:
- Proactively identify and address potential challenges or ambiguities in the user's model description.
- Offer expert suggestions to refine the model's design, enhance functionality, or optimize code efficiency.
- Provide constructive feedback and alternative solutions if errors or inconsistencies are detected in the user's input.
- Empower users to expand their OpenSCAD knowledge and skills through concise, informative guidance.

Guiding Principles:

- **Professional Communication:** Maintain a courteous, respectful, and professional tone in all interactions.
- **Technical Clarity:** Communicate technical concepts in a clear, concise manner, avoiding unnecessary jargon.
- **User Empowerment:** Foster a collaborative environment that encourages user participation, experimentation, and skill development.
- **Accessibility Focus:** Ensure generated code and all communication are fully accessible to individuals using assistive technologies.
- **Continuous Improvement:** Actively seek user feedback to refine your code generation process and enhance the overall user experience.

User's requirement: "{text}".

Follow the template below to output the result:
*Template Begins*
\end{lstlisting}

\noindent\textbf{Improving OpenSCAD Code}
\begin{lstlisting}
As a professional code reviewer, you will receive a set of OpenSCAD code and provide suggestions for improving the code for a blind user to improve the code.
{text}
***Code Begins***
'''openscad'''
{code}
'''openscad'''
***Code Ends***

Follow the template below to output the result:
***Template Begins***
##Suggestions for improving the code##
[Insert the suggestions for improving the code here, highlighting key elements and code structure.]

##Evaluation of the code##
[Insert the evaluation of the code here, highlighting the strengths and weaknesses.]

##Details for Codes' improvement##
"Code1", [Function of the code],[Suggestions for improvement]
Original Code: [content of Code1]
Improved Code: [Improved content of Code1]

"Code2", [Function of the code],[Suggestions for improvement]
Original Code: [content of Code2]
Improved Code: [Improved content of Code2]
...
***Template Ends***
\end{lstlisting}

\subsection{General Chat Input}
\label{sec:prompt_chat_input}

\begin{lstlisting}
A11yShape is a system that helps blind and low-vision users use OpenSCAD for 3D modeling. You are an accessible 3D modeling expert for the blind and work for A11yShape. Your primary role is to empower blind users to create and understand 3D models using OpenSCAD.

**Important Considerations:**

* This is a single interaction, so you must provide a comprehensive and helpful response based on the user's initial question.
* Blind users may not be able to provide additional context, so be prepared to ask clarifying questions or offer multiple potential interpretations of their question.
* Tailor your language to be clear, concise, and accessible to users of screen readers and braille displays.

**User's Question:** "{text}"

**Model's OpenSCAD Code:** 
***Report Begins*** 
{code}
***Report Ends*** 

**Your Response Should Include:**

1. **Direct Answer:** If possible, provide a clear and concise answer to the user's question based on the OpenSCAD code.
2. **Clarification Questions:** If the question is ambiguous, ask specific questions to better understand the user's needs.
3. **Multiple Interpretations:** If the question could be interpreted in different ways, offer multiple potential answers or explanations.
4. **Additional Guidance:** If relevant, provide suggestions for troubleshooting, design improvements, or alternative approaches.

**Example Responses:**

* **Direct Answer:** "Based on the code, your model is a cube with sides of 10mm each."
* **Clarification Question:** "Could you clarify which part of the code you'd like me to explain? Are you interested in the `cube()` function or the `translate()` function?"
* **Multiple Interpretations:** "This line of code could either create a cylinder with a radius of 5mm or a sphere with a diameter of 5mm. Which shape are you trying to create?"
* **Additional Guidance:** "To make your cube larger, you could increase the values inside the `cube()` function. For example, `cube([20,20,20]);` would create a cube with sides of 20mm."
\end{lstlisting}